\documentclass[lettersize,journal]{IEEEtran}
\usepackage{amsmath,amsfonts}
\usepackage{color}
\usepackage[noend]{algpseudocode}
\usepackage{amssymb}
\usepackage{array}
\usepackage{booktabs}
\usepackage[caption=false,font=footnotesize,labelfont=rm,textfont=rm]{subfig}
\usepackage{textcomp}
\usepackage{stfloats}
\usepackage{url}
\usepackage{cuted}
\usepackage{verbatim}
\usepackage{graphicx}
\usepackage{cite}
\usepackage{algorithm}
\usepackage{titlesec}
\usepackage{multicol}
\usepackage{algpseudocode}
\usepackage{etoolbox}
\usepackage{enumitem}
\setlist[itemize]{topsep=2pt,itemsep=1pt,parsep=0pt,partopsep=0pt}
\begin{document}

\pagestyle{empty}

\thispagestyle{empty}

\title{\huge{ 6DMA-Assisted Secure Wireless Communications}}

\author{Yanzhi Qian, 
Jing Jiang,~\IEEEmembership{Member,~IEEE},
Jingze Ding,~\IEEEmembership{Graduate Student Member,~IEEE}, 
Xiaodan Shao,~\IEEEmembership{Member,~IEEE}, 
Hongyun Chu,~\IEEEmembership{Member,~IEEE}

\thanks{This work was
supported in part by the National Natural Science Foundation of China under Grants 62371392, Grant 62201451 and Grant 62401643;
in part by the Key Program for Key
Industrial Chain Projects of Shaanxi Province under Grant 2023-ZDLGY-49 and Grant 2024GX-ZDCYL-05-01; Shaanxi Qinchuangyuan "Scientist and Engineer" Team Building Project 2024QCY-KXG-156 and Xi'an Science and Technology Plan Project under Grant 2023JH-GXRC-0180 and 2024JH-ZCLGG-0051; in part by the China Postdoctoral Science Foundation under Grant 2024M752439. (Corresponding author: Jing Jiang).}   

\thanks{Yanzhi Qian,  Jing Jiang, and Hongyun Chu are with the School of Communication and Information Engineering, Xi’an University of Posts and Telecommunications, Xi’an 710121, China (e-mail:qianstudent24@163.com; jiangjing@xupt.edu.cn; hychu@xupt.edu.cn).}

\thanks{Jingze Ding is with the School of Electronics, Peking University, Beijing 100871, China (e-mail: djz@stu.pku.edu.cn).
}

\thanks{X. Shao is with the Department of Electrical
and Computer Engineering, University of Waterloo, Waterloo, ON N2L 3G1, Canada (E-mail: x6shao@uwaterloo.ca).}

}
\markboth{}%
{}
\twocolumn
\maketitle 

\IEEEpubid{}

\maketitle
\thispagestyle{empty}
\begin{abstract}
Six-dimensional movable antenna (6DMA) has been widely studied for capacity enhancement, but its potential for physical layer security (PLS) remains largely unexplored. By adjusting both three-dimensional (3D) positions and 3D rotations of distributed antenna surfaces, 6DMA can increase spatial degrees of freedom (DoFs). The extra DoFs enable dynamic shaping of legitimate channels and suppresses eavesdropping channels, thereby offering unique advantages in enhancing secrecy performance.  Motivated by this, this letter proposes a novel 6DMA-assisted  secure wireless communication system, where the base station (BS) is equipped with 6DMA to enhance secrecy performance. Specifically, to simultaneously serve multiple legitimate users and counter cooperative interception by multiple eavesdroppers (Eves),  we formulate a sum secrecy rate (SSR) maximization problem by jointly optimizing the transmit and artificial noise (AN) beamformers, as well as the 3D positions and 3D rotations of antenna surfaces. To solve this non-convex problem, we propose an alternating optimization (AO) algorithm that decomposes the original problem into two subproblems and solves them iteratively to obtain a high-quality suboptimal solution. Simulation results demonstrate the superior secrecy performance over partially movable and conventional fixed-position antenna systems.

\end{abstract}

\begin{IEEEkeywords}
6DMA, physical layer security, artificial noise, antenna position and rotation optimization.
\end{IEEEkeywords}

\section{Introduction}
\IEEEPARstart{W}{ITH} the explosive growth of data traffic and the expansion of coverage, next-generation wireless communications are increasingly exposed to potential security risks. Therefore, physical layer security (PLS) that exploits channel randomness has attracted wide attention \cite{ref1}. Among various PLS approaches, secure beamforming is a key technique that enhances signals at legitimate users while suppressing those at eavesdroppers (Eves), the effectiveness is strongly influenced by the spatial degrees of freedom (DoFs) in multi-antenna systems \cite{refl}. However, conventional systems with fixed-position antennas (FPAs) have limited spatial DoFs and flexibility. As a result, their ability to adapt to dynamic wireless environments is restricted, thus constraining overall security performance. 

To overcome this limitation, 
movable antennas (MAs) have been explored in recent years to enhance PLS by dynamically adjusting antenna positions. Early studies focused on jointly optimizing antenna positions and transmit beamformers to maximize sum secrecy rate (SSR) and reduce power consumption \cite{ref2, ref3}. Extensions considered imperfect or statistical channel state information (CSI) \cite{ref4, ref5}, distributed access points \cite{ref6}, and full-duplex base station (BS) \cite{reff}, further demonstrating the potential of MAs in secure communications. 
However, existing MAs are restricted to one-dimensional or planar position motion with fixed antenna orientations, which limits their flexibility and adaptability to dynamic wireless environments. To fully exploit spatial channel variations, the concept of six-dimensional movable antennas (6DMA) was recently proposed \cite{ref7, ref8}. A 6DMA system comprises multiple antenna surfaces, each capable of 3D  translation and rotation,  thereby providing an innovative solution for highly flexible and adaptive antenna configurations. Furthermore, leveraging directional sparsity, channel estimation for all possible antenna positions and orientations was developed in \cite{ref9}. Compared with MA architectures, 6DMA provides unprecedented DoFs in spatial tunability, enabling antennas to flexibly avoid eavesdropping paths and reduce information leakage. Despite its promising advantages and potential performance gains, research on 6DMA-assisted secure communication systems remains largely unexplored.

Motivated by the above considerations, this letter investigates a 6DMA-assisted secure communication system, which consists of multiple  users, multiple Eves, and a  BS equipped with several 6DMA surfaces. To enhance security performance, an SSR maximization problem is formulated through the joint optimization of the 3D positions and 3D rotations of all 6DMA surfaces, as well as the transmit and artificial noise (AN) beamformers at the BS. To solve this non-convex optimization problem involving highly coupled variables, a modified alternating optimization (AO) algorithm is proposed, which decomposes the original problem into two subproblems and solves them iteratively. Simulation results validate the effectiveness of the proposed scheme under various configurations and demonstrate its superiority over conventional FPA systems and 6DMA systems with limited or partial movability.

\section{System Model}

\subsection{Channel Model}
\begin{figure}[t]
    \centering
\includegraphics[width=1\linewidth]{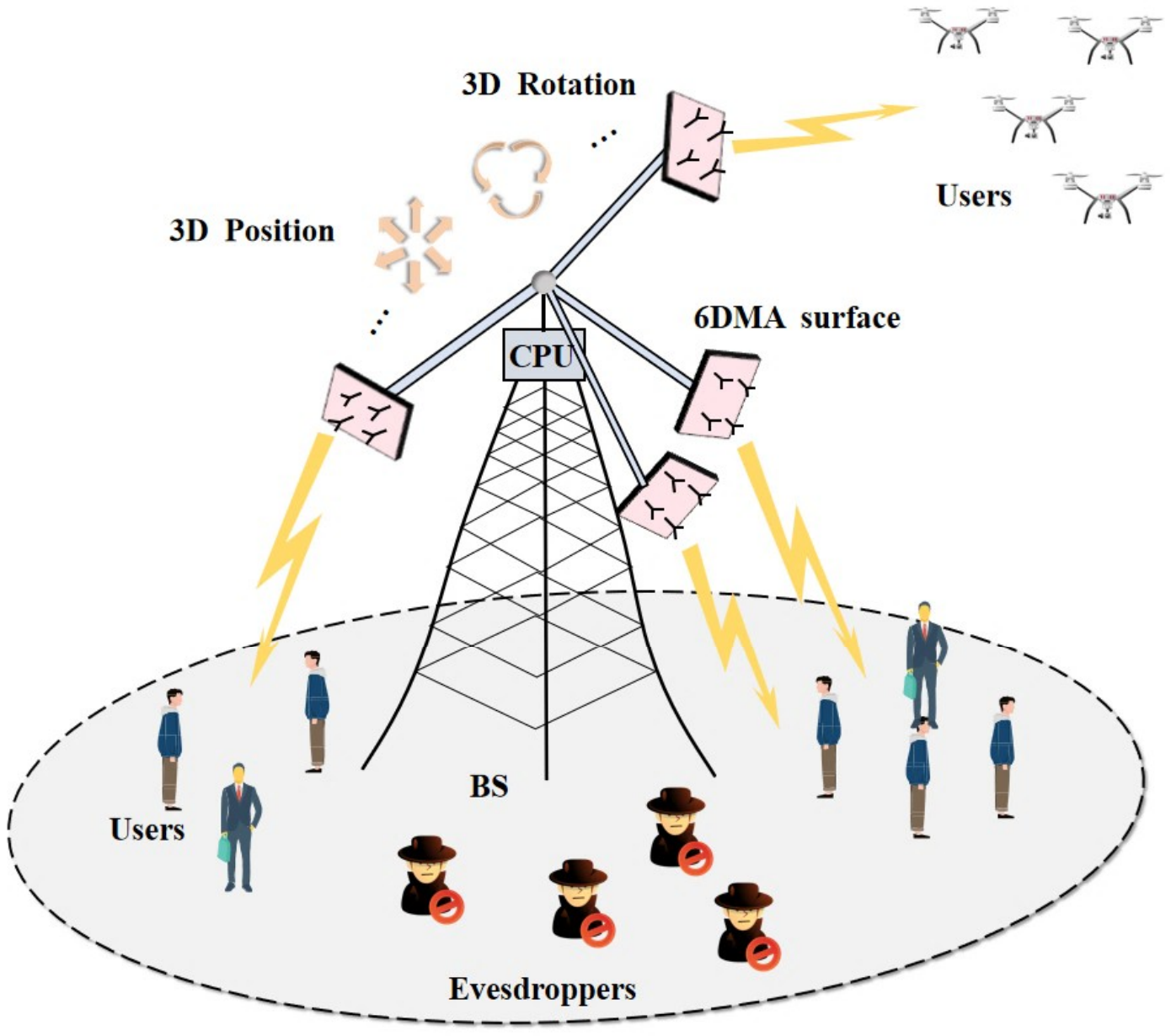}
    \caption{6DMA-aided secure communication.}
    \label{fig1}
\end{figure}
As shown in  Fig. \ref{fig1},  the  6DMA-assisted secure communication system  consists of a  BS, $K_D$  legitimate users, and $K_E$ Eves. Each user or Eve is equipped with a single FPA, while the BS employs $B$ 6DMA surfaces. Each 6DMA surface is assumed to be a uniform planar array (UPA)  containing $N$ antennas elements indexed by $\mathcal{N}=\left\{ 1,2,\ldots,N\right\}$. These 6DMA surfaces are mounted on extendable and rotatable rods with embedded flexible connections to the central processing unit (CPU), enabling joint adjustment of their 3D positions and 3D rotations within the deployment region
 to positively reconfigure the channel conditions. Specifically, the position and rotation of the $b$-th 6DMA surface ($b \in \mathcal{B}=\left\{ 1,2,\ldots,B\right\}$) can be written as

\begin{equation}
\label{deqn_ex1a}
{\mathbf{q}_{b}}= \left[ x_{b},y_{b},z_{b} \right]^{T} \in \mathbb{R}^{3 \times 1},
\end{equation}
\begin{equation}
\label{deqn_ex1a}
\begin{aligned}
{\mathbf{u}_{b}= \left[ \alpha_{b}, \beta_{b}, \gamma_{b} \right]^{T} \in \mathbb{R}^{3 \times 1}},
\end{aligned}
\end{equation}
where $x_{b}$, $y_{b}$ and $z_{b}$ denote the coordinates of the $b$-th 6DMA surface's center in the global Cartesian coordinate system (CCS), with the CPU of the BS taken as the reference origin. The variables $\alpha_{b} \in [0, 2\pi)$, $\beta_{b} \in [0, 2\pi)$, and $\gamma_{b} \in [0, 2\pi)$ represent the Euler angles
corresponding to rotations around the $x$-, $y$-, and $z$-axes, respectively. Accordingly, the global position of the $n$-th antenna on the $b$-th 6DMA surface is given by

\begin{equation}
{\mathbf{r}}_{b,n}\big({\mathbf{q}}_{b},{\mathbf{u}}_{b}\big) = {\mathbf{q}}_{b}+{\mathbf{R}}\big({\mathbf{u}}_{b}\big){ \mathbf{\overline{r}}}_{n},n\in\mathcal{N},b\in\mathcal{B},
\end{equation}
where the rotation matrix $\mathbf{R}(\mathbf{u}_{b})$ is constructed based on ${\mathbf{u}}_{b}$, and $\mathbf{\overline{r}}_{n}$ denotes the position of the $n$-th antenna on the 6DMA surface in its local CCS\cite{ref7}. \\
\indent We neglect the non-line-of-sight (NLoS) components and only consider the dominant line-of-sight (LoS) channel between each legitimate user or Eve and the BS. Let $\phi_{b} \in[- \pi, \pi]$ and $\theta_{b}\in \left[-\pi / 2,\pi / 2\right]$ represent the azimuth and elevation angles, respectively, of a signal transmitted to any user or Eve with respect to the center of the $b$-th surface. The unit-length direction-of-departure (DOD) vector $\mathbf{f}_{b}$ corresponding to direction $(\phi_{b},\theta_{b})$ is given by

\begin{equation}
\mathbf{f}_{b}= \left[ \cos \left( \theta_{b} \right) \cos \left( \phi_{b} \right), \cos \left( \theta_{b} \right) \sin \left( \phi_{b} \right), \sin \left( \theta_{b} \right) \right]^{\text{T}}.
\end{equation}
\indent Based on the above, the hybrid-field channel model between legitimate user or Eve $k$ and the BS can be expressed as \cite{ref10}
\begin{align}
\mathbf{h}_{{}_{k}}(\mathbf{q},\mathbf{u})
&=v[  \sqrt{g(\mathbf{q}_{{}_{1}}, \mathbf{u}_{{}_{1}})} e^{-j\frac{2\pi}{\lambda}d_{\mathrm{1}}} 
\mathbf{a}( \mathbf{q}_{{}_{1}},\mathbf{u}_{{}_{1}})^{{}^{\text{T}}}, \notag \\
&\quad \cdots,  \sqrt{g(\mathbf{q}_{{}_{B}},\mathbf{u}_{{}_{B}})} e^{-j\frac{2\pi}{\lambda}d_{{}_{B}}} 
\mathbf{a}(\mathbf{q}_{{}_{B}},\mathbf{u}_{{}_{B}})^{{}^{\text{T}}} \big]\in\mathbb{C}^{1\times NB}  ,\label{h_vector}
\end{align}
where $\mathbf{q}=[\mathbf{q}_{1},\mathbf{q}_{2},\cdots,\mathbf{q}_{B}]$ and $\mathbf{u}=[\mathbf{u}_{1},\mathbf{u}_{2},\cdots,\mathbf{u}_{B}]$. Here, $v$ denotes the path gain, and $d_b$ denotes the distance from user or Eve $k$ to the center of the $b$-th surface . 
The index $k$ belongs to either the user set $\mathcal{K}_D = \{1, \ldots, K_D\}$ or the Eve set $\mathcal{K}_E = \{1, \ldots, K_E\}$. Consequently, the steering vector of the $b$-th 6DMA surface $\mathbf{a}(\mathbf{q}_{ {}_{b}},\mathbf{u}_{ {}_{b}})\in\mathbb{C}^{{N}\times1}$ can be express as

\begin{equation}
\mathbf{a}\left(\mathbf{q}_{{}_{b}},\mathbf{u}_{{}_{b}}\right) = \left[ e^{j \frac{2 \pi}{\lambda}\mathbf{f}_{b}^{T} \mathbf{r}_{b,1}\left(\mathbf{q}_{{}_{b}},\mathbf{u}_{{}_{b}}\right)},\,\cdots,\,e^{j \frac{2 \pi}{\lambda} \mathbf{f}_{b}^{T} \mathbf{r}_{b,N}\left(\mathbf{q}_{{}_{b}},\mathbf{u}_{{}_{b}}\right)} \right]^{\text{T}},
\label{eq:steering_vector}
\end{equation}
where ${\lambda}$ denotes the carrier wavelength. In \eqref{h_vector}, the antenna gain is given by  $g \left( \mathbf{q}_{b},\mathbf{u}_{b} \right)=10^{ \frac{A \left( \widetilde{ \theta}_{b}, \widetilde{ \phi}_{b} \right)}{10}}$, where the angles $\widetilde{ \theta}_{b}$ and $\widetilde{ \phi}_{b}$ can be projected by $\mathbf{f}_{b}$, and $A \left( \widetilde{ \theta}_{b}, \widetilde{ \phi}_{b} \right)$ denotes  the effective antenna gain (in dBi) of the $b$-th 6DMA  surface\cite{ref7}.   

\subsection{Signal Model}
\indent  In each time slot, the BS transmits \( K_{D} \) independent
information streams to \( K_{D} \) users, while an AN vector \( \mathbf{v} \in \mathbb{C}^{NB \times 1} \) is simultaneously transmitted to degrade Eves' malicious interception. Therefore, the downlink signals is expressed as
\begin{equation}
    \mathbf{x} = {\mathbf{W}}\mathbf{s} + \mathbf{v}, 
    \label{eq:transmit_signal}
\end{equation}
where ${\mathbf{W}} = [\mathbf{w}_1, \cdots, \mathbf{w}_{K_D}] \in \mathbb{C}^{NB \times K_D}$ denotes the transmit beamforming matrix, 
$\mathbf{s} = [s_1, \cdots, s_{K_D}]^T \in \mathbb{C}^{K_D \times 1}$ 
represents the transmitted symbols, 
with $\mathbb{E}\{\mathbf{s}\mathbf{s}^H\} = \mathbf{I}_{K_D}$. For user $k_d\in\mathcal{K}_D$ and Eve $k_e\in\mathcal{K}_E$, the received signals are
respectively given by  
 
\begin{align}
y_{{}_{k_d}} 
&= \underbrace{\mathbf{h}_{{}_{k_d}}\left(\mathbf{q,u}\right)\mathbf{w}_{{}_{k_d}}{s}_{{}_{k_d}}}_{{\text{downlink desired information}}} 
+ \underbrace{\sum_{{}_{i\in\mathcal{K}_{\mathrm{D}}} \setminus \left\{k_d\right\}} \mathbf{h}_{{}_{k_d}}\left(\mathbf{q,u}\right) \mathbf{w}_{{}_{i}} {s}_{{}_{i}}}_{{\text{downlink multi-user interference}}} \notag \\
&\quad + \underbrace{\mathbf{h}_{{}_{k_d}}\left(\mathbf{q,u}\right)\mathbf{v}}_{{\text{AN-interference}}} + {n}_{{}_{k_d}} ,\label{eq:ykd_rx}
\end{align}
\begin{align}
y_{{}_{k_e}} 
&= \sum_{k_d \in \mathcal{K}_{\mathrm{D}}} 
\underbrace{\mathbf{h}_{{}_{k_e}}\left(\mathbf{q}, \mathbf{u}\right) \mathbf{w}_{{}_{k_d}} {s}_{{}_{k_d}}}_{\text{downlink information}} 
+ \underbrace{\mathbf{h}_{{}_{k_e}}\left(\mathbf{q}, \mathbf{u}\right)\mathbf{v}}_{\text{AN-interference}}  +{n}_{{}_{k_e}}, \label{eq:yke_rx}
\end{align}
where ${n}_{k_d}\sim\mathcal{C}N\left(0,\sigma_{k_d}^{2} \right)$ and ${n}_{k_e}\sim\mathcal{C}N\left(0,\sigma_{k_e}^{2} \right)$ denote the additive white Gaussian noise (AWGN) at the user $k_d$ and Eve $k_e$, respectively. To ensure secure communication, this letter assumes a worst-case scenario in which all Eves jointly process the received confidential information and are able to completely eliminate multi-user interference before decoding \cite{ref}. Under these assumptions, the achievable rates of the legitimate user $k_d$ and Eve $k_e$ are given by \eqref{eq:10} and \eqref{eq:11} at the bottom of next page, where $\gamma _{k_d}$ denotes the signal-to-interference-noise ratio (SINR) at user $k_d$ and $\gamma _{k_d}^{k_e}$ represents the SINR of Eve $k_e$ eavesdropping on the user $k_d$.   

\begin{figure*}[b]  
\hrulefill  
\begin{equation}
\label{eq:10}
R_{k_d} = \log_2(1 + \gamma_{k_d}) = \log_2\left(1 + \frac{ \left| \mathbf{h}_{k_d}(\mathbf{q}, \mathbf{u}) \mathbf{w}_{k_d} \right|^2 }{ \sum\limits_{i \in \mathcal{K}_D \setminus \{k_d\}} \left| \mathbf{h}_{k_d}(\mathbf{q}, \mathbf{u}) \mathbf{w}_{i} \right|^2 + \left| \mathbf{h}_{k_d}(\mathbf{q}, \mathbf{u}) \mathbf{v} \right|^2 + \sigma_{k_d}^2 } \right).
\end{equation}

\vspace*{-1em} 
\end{figure*}

\begin{figure*}[b]
\begin{equation}
\label{eq:11}
R_{k_d}^{E} = \log\left(1 + \sum_{k_e \in \mathcal{K}_E} \gamma_{k_d}^{k_e} \right) = \log\left( 1 + \sum_{k_e \in \mathcal{K}_E} \frac{ \left| \mathbf{h}_{k_e}(\mathbf{q}, \mathbf{u}) \mathbf{w}_{k_d} \right|^2 }{ \left| \mathbf{h}_{k_e}(\mathbf{q}, \mathbf{u}) \mathbf{v} \right|^2 + \sigma_{k_e}^2 } \right).   
\end{equation}
\vspace*{-0.5em}
\end{figure*}
\subsection{Problem Formulation}
In this letter, we aim to maximize the SSR of all users 
\begin{equation}
R_\text{SSR} = \sum\limits_{k_d \in \mathcal{K}_{\mathrm{D}}} \left[R_{k_d} - R_{{k_d}}^{E}\right]^{+},  \end{equation}
by jointly optimizing the 3D positions $\mathbf{q}$ and 3D rotations $\mathbf{u}$  of all 6DMA surfaces, as well as the transmit beamformer $\mathbf{w}$ and AN beamformer $\mathbf{v}$ at the BS.  Assume perfect CSI is available at the BS. The corresponding optimization problem is formulated as follows.    

\begin{align}
\text{(P1)}: \quad &\mathop{\mathrm{max}}_{\mathbf{q}, \mathbf{u}, \mathbf{W}, \mathbf{v}} \quad R_{\text{SSR}} \\
\text{s.t.} \quad
&\text{C1: } \mathbf{q}_i \in \mathcal{C}, \quad \forall i \in \mathcal{B}, \notag \\
&\text{C2: } \sum\limits_{k_d \in \mathcal{K}_{\mathrm{D}}}  \mathbf{w}_{k_d}^{H} \mathbf{w}_{k_d} 
+  \mathbf{v}^{H} \mathbf{v} \leq P_{\max}, \notag \\
&\text{C3: } \left\| \mathbf{q}_i - \mathbf{q}_j \right\|_2 \geq d_{\min}, \quad i,j \in \mathcal{B},\ j \ne i, \notag \\
&\text{C4: } \mathbf{n}(\mathbf{u}_i)^{T} (\mathbf{q}_j - \mathbf{q}_i) \leq 0, \quad i,j \in \mathcal{B},\ j \ne i, \notag \\
&\text{C5: } \mathbf{n}(\mathbf{u}_i)^{T} \mathbf{q}_i \geq 0, \quad i \in \mathcal{B}, \notag
\end{align}
where $\mathbf{n}(\mathbf{u}_b)$ denotes the normal vector of the $b$-th 6DMA surface in the local CCS. Constraint C1 ensures that the movement regions of all 6DMA surfaces are confined within the given 3D deployment space $\mathcal{C}$. Constraint C2 limits the maximum downlink transmit power $P_{\max}$ of the BS. Constraint C3 imposes a minimum separation $d_\text{min}$ between 6DMA surfaces to eliminate overlap and electromagnetic coupling, while constraint C4 excludes mutual reflections among antennas and constraint C5 mitigates blockage effects from the BS's CPU. The operator $[x]^+ = \max\{x,0\}$ does not affect the optimization and thus is omitted in the subsequent analysis. Owing to the strong interdependence among variables together with the non-convex nature of both the objective and the constraints, problem (P1) is difficult to solve directly; hence, an AO-based algorithm is developed in the following section.

\section{Proposed Solution}

In this section, instead of jointly optimizing $\mathbf{q}$, $\mathbf{u}$,  $\mathbf{W}$, and $\mathbf{v}$ with high computational complexity, an AO algorithm is proposed to solve problem (P1). The optimization procedure is divided into two subproblems. First, given fixed transmit and AN beamformers, the 3D positions $\mathbf{q}$ and 3D rotations $\mathbf{u}$ of all 6DMA surfaces are optimized by using the penalty successive convex approximation (PSCA) method. Then, with the antenna positions and rotations fixed, the transmit beamformer $\mathbf{W}$ and AN beamformer $\mathbf{v}$ are subsequently updated.

\subsection{Sub-Problem 1: Optimization of $\{\mathbf{q}, \mathbf{u}\}$ with Given $\{\mathbf{W}, \mathbf{v}\}$}
At each iteration of the proposed AO algorithm, the variables ${\mathbf{q}, \mathbf{u}}$ are optimized by sequentially updating the position and rotation of each 6DMA surface (i.e., $\mathbf{q}_b$ and $\mathbf{u}_b$, $\forall b \in \mathcal{B}$) while keeping the others fixed. 

Given $\mathbf{u}_{b}$, constraints C4 and C5 are affine with respect to $\mathbf{q}$ and thus remain unchanged.  The non-convex constraint C3 is linearized by applying the first-order Taylor expansion around the value of $\mathbf{q}_b$ in the previous $(t -1)$-th iteration, leading to the following convex inner approximation:
\begin{equation}
-\|\mathbf{q}_b^{t-1} - \mathbf{q}_j\|^2 
- 2(\mathbf{q}_b^{t-1} - \mathbf{q}_j)^\top (\mathbf{q}_b - \mathbf{q}_b^{t-1})
\le -d_{\min}^2.
\label{eq:sca_min_dist}
\end{equation}
\indent Thus, the update $\mathbf{q}_b^t$ is obtained by solving the following convex quadratic sub-problem:
\begin{align}
\text{(P2)}: \quad 
&-\min_{\mathbf{q}_{b}} \ \nabla R_{\mathrm{SSR}}\left(\mathbf{q}_b^{t-1}\right)^{T} \left(\mathbf{q}_{b}-\mathbf{q}_{b}^{t-1}\right) \notag \\
&\quad\quad + \frac{\rho}{2}\|\mathbf{q}_{b}-\mathbf{q}_{b}^{t-1}\|^{2} \notag \\
&\text{s.t. } \text{C1},\ \eqref{eq:sca_min_dist},\ \text{C4},\ \text{C5}, \notag
\end{align}
where $\rho > 0$ is the proximal parameter that guarantees strong convexity and stabilizes the iteration. Here, $\nabla R_{\mathrm{SSR}}$ is approximated by its first-order expansion around $\mathbf{q}_{b}^{t-1}$, which can be given by 
\begin{align}
\label{eq:15}
&\left[\nabla R_{\mathrm{SSR}}\left( \mathbf{q}_{b}^{t-1}\right) \right]_{j}\\
&= \lim_{\varepsilon \to 0} \frac{1}{\varepsilon} \Big( 
R_{\mathit{SSR}} \left( \mathbf{q}_{b}^{t-1} + \varepsilon \mathbf{e}^{j} \right) - R_{\mathit{SSR}} \left( \mathbf{q}_{b}^{t-1} \right) \Big), \notag
\end{align}
where $\mathbf{e}^{j}\in\mathbb{R}^{3\times 1}$ is a vector whose $j$-th element is 1 and the remaining elements are 0.  Note that problem (P2) is a convex quadratic optimization problem that can be efficiently solved using standard solvers (e.g., \texttt{quadprog} in MATLAB).   

Given $\mathbf{q}_b$, $\mathbf{u}_b$ can be optimized in the similar way. However, constraints C4 and C5 become non-convex due to their dependence on $\mathbf{n}(\mathbf{u}_b)$. Fortunately, they can be relaxed into convex ones by linearly approximating the rotation matrix $\mathbf{R}\left(\mathbf{u_{b}}\right)$, which is specified in (52)-(58) of \cite{ref7}. Thus, the update $\mathbf{u}_b^t$ is obtained by solving the following convex quadratic subproblem:
\begin{align}
\text{(P3)}: \quad 
&-\min_{\mathbf{u}_{b}} \ \nabla R_{\mathrm{SSR}}\left(\mathbf{u}_b^{t-1}\right)^{T} \left(\mathbf{u}_{b}-\mathbf{u}_{b}^{t-1}\right) \notag \\
&\quad\quad +\frac{\rho}{2}\|\mathbf{u}_{b}-\mathbf{u}_{b}^{t-1}\|^{2} \notag \\
&\text{s.t. }\ \text{(57)},\ \text{(58) of [9]}. \notag
\end{align}
\subsection{Sub-Problem 2: Optimization of $\{\mathbf{W}, \mathbf{v}\}$ with Given $\{\mathbf{q},\mathbf{u}\}$}

 To mitigate the strong coupling between $\mathbf{W}$ and $\mathbf{v}$ arising from their shared power budget in constraint C2, we introduce an auxiliary power allocation factor $\alpha \in (0,1)$ to decouple them. $\alpha P_{\max}$ of the power budget is allocated to the information beamformer $\mathbf{W}$, while the remaining $(1-\alpha)P_{\max}$ is reserved for the AN vector $\mathbf{v}$. 

  For any given positions $\mathbf{q}$ and rotations $\mathbf{u}$ of the 6DMA surfaces, the transmit beamformer $\mathbf{W}$ at the BS is updated under the minimum mean square error (MMSE) criterion, i.e.,
\begin{equation}
\mathbf{W} = \left( \mathbf{H}(\mathbf{q}, \mathbf{u})^{\mathrm{H}} \mathbf{P} \mathbf{H}(\mathbf{q}, \mathbf{u}) + \sigma^2 \mathbf{I}_{NB} \right)^{-1} \mathbf{H}(\mathbf{q}, \mathbf{u})^{\mathrm{H}},     
\label{eq:16}
\end{equation}
where $\mathbf{H}(\mathbf{q}, \mathbf{u})=\left[ \mathbf{h}_1(\mathbf{q}, \mathbf{u}); \ldots; \mathbf{h}_{K_D}(\mathbf{q}, \mathbf{u})\right]\in\mathbb{C}^{K_{D}\times NB}$ denotes the downlink channel matrix from the BS to the $K_D$ users. The diagonal matrix $\mathbf{P} =\mathrm{diag}(P_1, \dots, P_{K_D})$ specifies the transmit power allocated to individual users, with each user $i$ allocated
 $P_i = \frac{\alpha P_{\max} \, \left\| \mathbf{h}_i(\mathbf{q},\mathbf{u}) \right\|^2}
{\sum\limits_{\ell \in \mathcal{K}_D} \left\| \mathbf{h}_\ell(\mathbf{q},\mathbf{u}) \right\|^2 }$.
 
To avoid interference with legitimate users, the AN beamformer $\mathbf{v}$ is projected onto the null space of $\mathbf{H}(\mathbf{q}, \mathbf{u})$. Specifically,  let $\mathbf{U}_{\text{null}}$ denotes an orthonormal basis of the null space of $\mathbf{H}(\mathbf{q}, \mathbf{u})$ obtained via singular value decomposition (SVD) \cite{ref11}. The downlink channel matrix between the BS and the $K_E$ Eves is denoted by $\mathbf{H}_{\text{eve}}(\mathbf{q}, \mathbf{u}) \in \mathbb{C}^{K_{E}\times NB}$, where each row corresponds to the channel vector of one Eve. The artificial noise (AN) vector is then given by
\begin{equation}
\mathbf{v} = \sqrt{(1-\alpha)P_{\max}} \cdot \mathbf{U}_{\text{null}}. \mathbf{z}_{\max},
\label{eq:17}
\end{equation}
with $\mathbf{z}_{\max} 
= \arg\max\limits_{\|\mathbf{z}\|=1} 
\mathbf{z}^{\text{H}}\mathbf{U}_{\text{null}}^{\text{H}}\mathbf{H}_{\mathrm{eve}}^{\text{H}}(\mathbf{q}, \mathbf{u})
\mathbf{H}_{\text{eve}}(\mathbf{q}, \mathbf{u})\mathbf{U}_{\text{null}}\mathbf{z}$. Here, $\mathbf{z}$ is a unit-norm vector in the null space of the legitimate users’ channel, and $\mathbf{z}_{\max}$ is its optimal choice, i.e., the dominant eigenvector that maximizes AN leakage to the eavesdroppers.  The design of $\mathbf{v}$ remains orthogonal to all legitimate users’ channels while maximizing the interference power at the Eves. Finally, a one-dimensional search over a discrete set of $\alpha$ values is conducted to identify the optimal power allocation. The procedure for addressing problem (P1) is outlined in Algorithm 1. \\
\indent The AO- and PSCA-based updates guarantee that the objective value of problem (P1) is monotonically non-decreasing across iterations, and the constrained formulation imposes a finite upper bound, thereby ensuring the convergence of Algorithm 1. As for the computational complexity, the majority of the computational cost arises from the iterative process in lines 5–15. Therefore,  the overall complexity of the proposed  algorithm is $\mathcal{O}(T_1 T_2 N B^{3} (K_D+K_E)^2)$.

\setlength{\textfloatsep}{6pt}
\setlength{\intextsep}{6pt}

\algtext{EndFor}{\textbf{end for}}
\algtext{EndWhile}{\textbf{end while}}
\algtext{EndIf}{\textbf{end if}}
\AtBeginEnvironment{algorithmic}{\setlength{\itemsep}{0pt}}
\begin{algorithm}
\caption{Proposed AO Algorithm for Solving (P1)}
\label{alg1}
\begin{algorithmic}[1]
\State \textbf{Input}: \(\{\mathbf{q}^0_b\}_{b=1}^B\), \(\{\mathbf{u}^0_b\}_{b=1}^B\), \(\alpha_{\min}\), \(\alpha_{\max}\), \(P_{\max}\), \(B\), \(S\), \(\delta\), and iteration numbers \(T_1\), \(T_2\).   
\For{\(m = 1:T_1\)}
    \State Update \(\mathbf{W}\) via \eqref{eq:16};
    \State \(t = 1\);
    \While{Increase of SSR is above \(\delta\) \textbf{or} \(t < T_2\)}
        \For{\(b = 1:B\)}
            \State Calculate \(\nabla_{\mathbf{q}_b} R_{\text{SSR}}(\mathbf{q}_b^{t-1})\) via \eqref{eq:15};
            \State Update \(\mathbf{q}_b^t \) by solving problem (P2);
        \EndFor
        \For{\(b = 1:B\)}
            \State Calculate \(\nabla_{\mathbf{u}_b} R_{\text{SSR}}(\mathbf{u}_b^{t-1})\) via (62) in [9];
            \State Update \(\mathbf{u}_b^t \) by solving problem (P3);
        \EndFor
        \State \(t = t + 1\);
    \EndWhile
    
    \State $\xi = -\infty$;
    \For{\(\alpha = \alpha_{\min}:S:\alpha_{\max}\)}
        \State Update \(\mathbf{W}\) via \eqref{eq:16};
        \State Update \(\mathbf{v}\) via \eqref{eq:17};
        \If{\(R_\text{SSR}(\mathbf{q}, \mathbf{u}, \mathbf{W}, \mathbf{v}) > 
        \xi\)}
            \State \( \xi = R_\text{SSR}(\mathbf{q}, \mathbf{u}, \mathbf{W}, \mathbf{v}) \);
            \State \(\mathbf{W}^{opt} = \mathbf{W}\);
            \State\(\mathbf{v}^{opt} = \mathbf{v}\);
        \EndIf
    \EndFor
\EndFor
\State \Return \(\mathbf{q}, \mathbf{u}, \mathbf{W}^{opt}, \mathbf{v}^{opt}\).
\end{algorithmic}
\end{algorithm}
\section{Simulation Results}

To evaluate the effectiveness of the proposed scheme, we consider the following simulation setup. The general non-homogeneous Poisson process (NHPP) model is adopted to generate the number and spatial distribution of $K_D$ users and $K_E$ Eves, where the probability mass functions of both $K_D$ and $K_E$ follow Poisson distributions \cite{ref12}. Unless otherwise indicated, the parameter settings are listed in Table I. In the simulations, the proposed scheme is referred to as ``Proposed''. In addition, to highlight the performance gain of the proposed scheme, the following benchmark strategies are considered.

\vspace{-1mm} 
\begin{itemize}
    \item \textbf{FPA}: In this scheme, each sector antenna covers approximately $120^{\circ}$, with all UPAs fixed in both position and orientation. The downtilt angle is statically set to $ 15^{\circ}$.
    
    \item \textbf{6DMA with circular movement}: The downtilt of each sector antenna is fixed at $15^{\circ}$. The antenna centers are allowed to move along a circular trajectory parallel to the ground plane.
    
    \item \textbf{6DMA with flexible rotation only}: The center positions of all sector antennas are fixed, while only their rotations are optimized by using the proposed algorithm.
\end{itemize}
\vspace{-1mm}
\begin{table}[!htbp] 
\centering 
\caption{Simulation Parameters} 
\label{table_example} 
\begin{tabular}{ll} 
\toprule 
\multicolumn{1}{m{5cm}}{\raggedright Parameters} &  
\multicolumn{1}{m{1cm}}{\raggedright Value}\\
\midrule 
User-BS distance $d$&20-200m\\ 
Number of channel paths $L$&1\\
Average noise power $\sigma_{k_d}^{2}$ and $\sigma_{k_e}^{2}$ &-90dBm\\
Maximum transmit power of the BS $P_{\max}$ &10W\\
Convergence thresholds $\delta$ &$10^{-3}$\\
Minimum antenna spacing ${{d}_{min}}$ &$\frac{ \sqrt{2}}{2} \lambda+ \frac{ \lambda}{4}$ \\
Carrier wave-length $\lambda$&0.125m\\
Number of users $K_D$&7\\
Number of Eves $K_E$ &1\\
Number of antennas of each 6DMA surface $N$&4\\
Number of surfaces $B$&8\\
\bottomrule
\end{tabular}
\end{table}

\begin{figure*}[t]
    \centering
    \begin{minipage}[t]{0.32\textwidth}
        \centering
        \includegraphics[width=\linewidth]{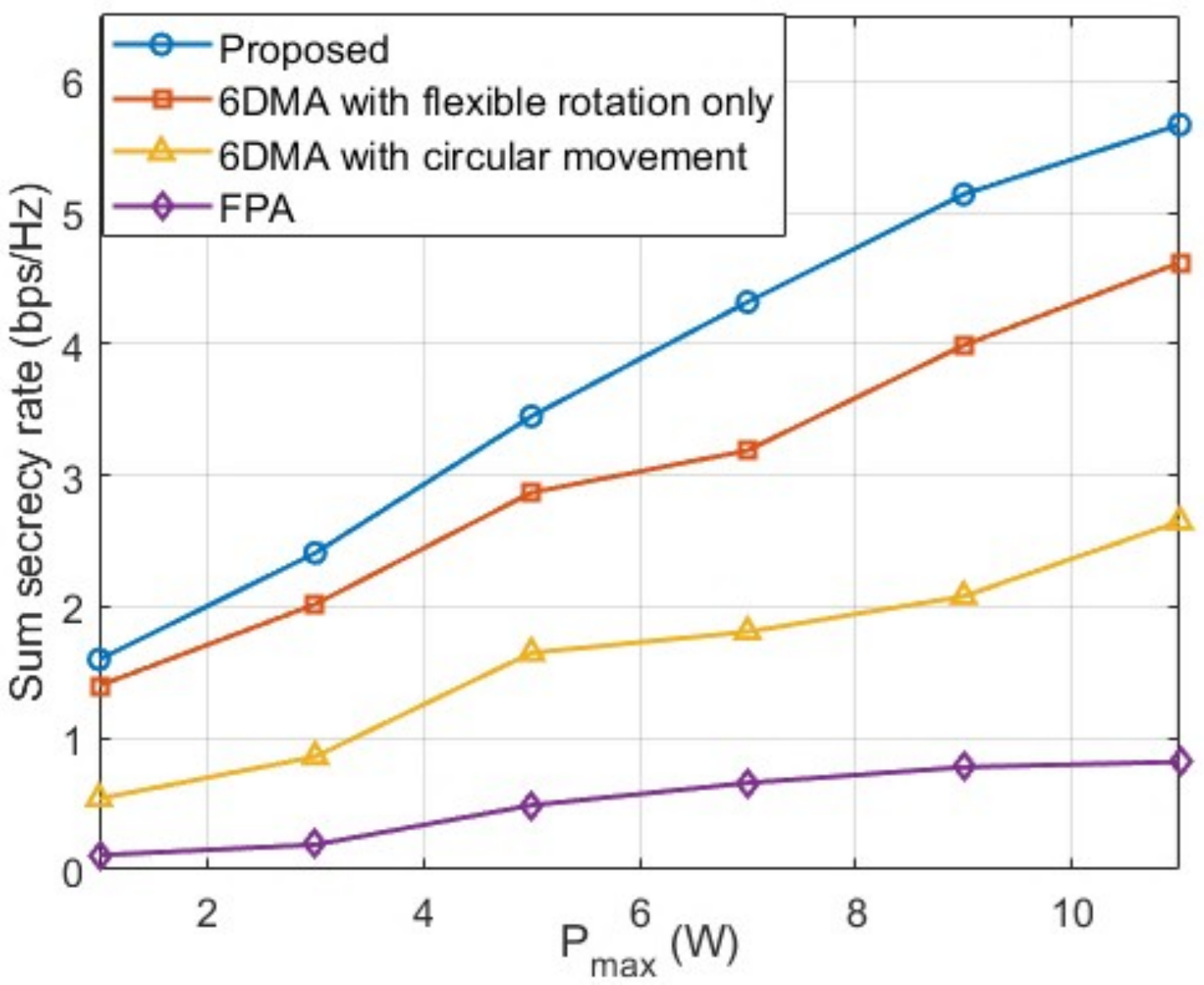}
        \caption{Sum secrecy rate versus BS transmit power in 3D user distribution scenarios.}
        \label{fig2}
    \end{minipage}
    \hfill
    \begin{minipage}[t]{0.32\textwidth}
        \centering
        \includegraphics[width=\linewidth]{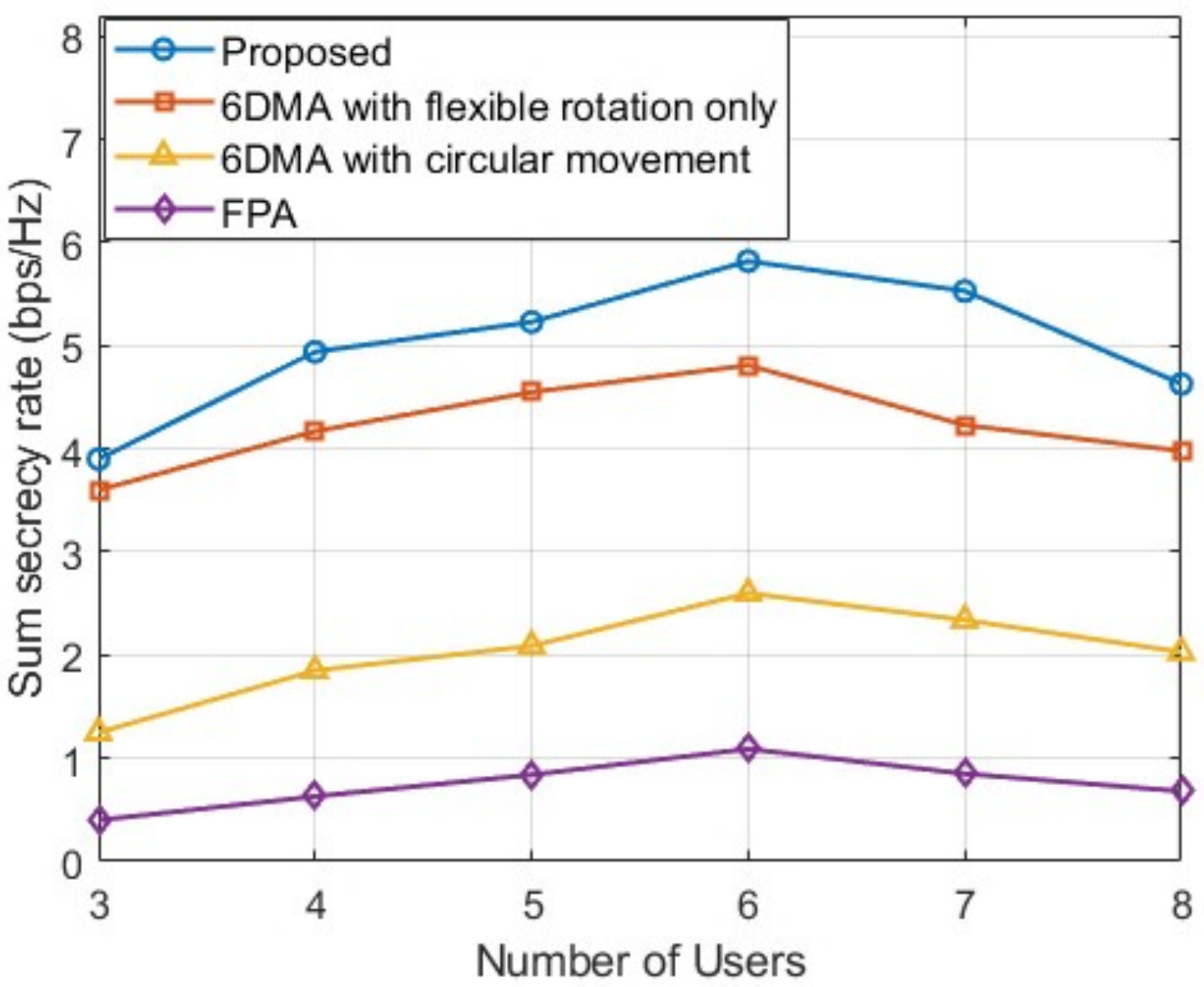}
        \caption{Sum secrecy rate versus average number of users in 3D user distribution scenarios.}
        \label{fig3}
    \end{minipage}
    \hfill
    \begin{minipage}[t]{0.32\textwidth}
        \centering
        \includegraphics[width=\linewidth]{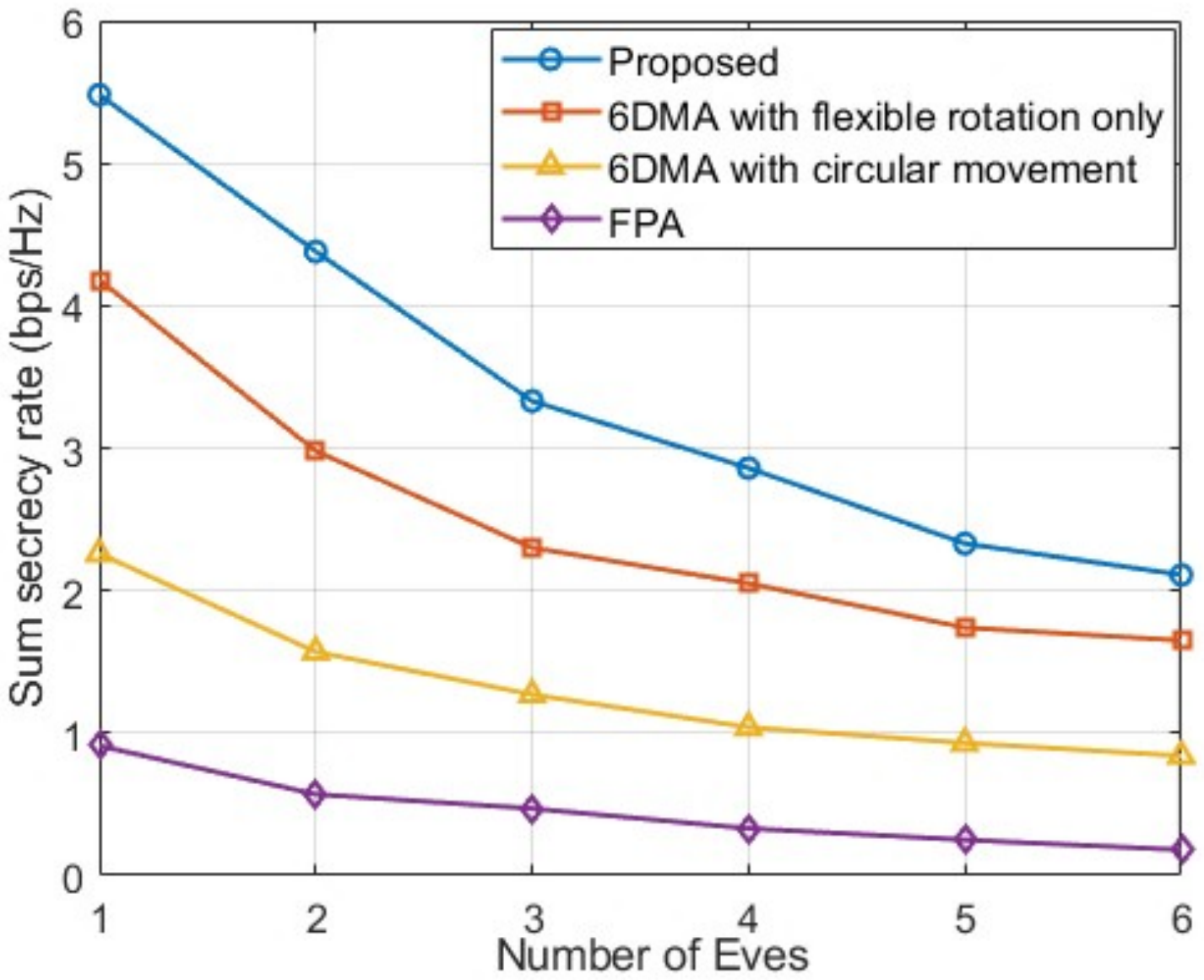}
        \caption{Sum secrecy rate versus average number of Eves in 3D user distribution scenarios.}
        \label{fig4}
    \end{minipage}
\end{figure*}

Fig. \ref{fig2} presents the SSR performance of the proposed and benchmark schemes versus the BS transmit power. It is observed that the SSR of all the schemes are significantly improved with the increase of the transmit power budget, where the proposed scheme achieves larger SSR performance. 
Moreover, the performance gap between the proposed and benchmark schemes widens as the transmit power increases. This is expected, as higher transmit power allows the proposed scheme to better leverage its spatial design flexibility and interference suppression capabilities, resulting in a growing performance advantage over the baseline schemes.

Fig. \ref{fig3} illustrates the trend of the SSR as a function of the number of users under different schemes. As expected, the SSR of all schemes initially increases with the number of users. When the number of users reaches 6, the proposed  scheme attains its peak performance, achieving gains of 1.5 dB, 3.9 dB, and 4.2 dB compared with the 6DMA with flexible rotation only, the 6DMA with circular movement, and the conventional FPA scheme, respectively. After this point, the SSR of all schemes gradually decreases as user interference begins to dominate. Nevertheless, the proposed scheme consistently outperforms the benchmarks. These improvements stem from the joint exploitation of both the positional and rotational DoFs of the 6DMA surfaces, which provides superior array gain and spatial multiplexing capability. In contrast, the FPA and limited-mobility schemes cannot fully utilize such flexibility, leading to less pronounced gains.

As illustrated in Fig. \ref{fig4}, the SSR of all considered schemes declines progressively with the increasing number of eavesdroppers. This is because a larger number of Eves enhances the aggregate interception capability of the wiretap channels, leading to a reduced SSR and weaker physical-layer security. Nevertheless, the proposed scheme still significantly outperforms the other benchmark schemes, thereby demonstrating its robustness in safeguarding confidential transmissions even in densely wiretapped environments.

\section{Conclusion}
In this letter, we investigated a novel 6DMA-assisted secure communication framework that leverages DoFs in both 3D positioning and rotation of the 6DMA surfaces. To enhance the achievable SSR, an efficient algorithm was proposed by jointly optimizing the positions and rotations of all 6DMA surfaces, as well as the transmit and AN beamformers at the BS. Numerical results were presented to evaluate the performance of the proposed scheme, demonstrating its superiority over the conventional FPA scheme and 6DMA schemes with limited or partial movability.

\newpage

\vfill
\end{document}